\documentclass{emulateapj}

\usepackage{graphics,graphicx}
\usepackage{natbib}
\usepackage{epstopdf}
\usepackage{color}
\def\apj{\rm ApJ}

\def\aap{\rm A\&A}
\newcommand{\msun}{M$_{\odot}$}

\shorttitle{Migration Traps in AGN Disks}
\shortauthors{Bellovary et al.}

\begin{document}

\title{Migration Traps in Disks Around Supermassive Black Holes}

\author{Jillian M. Bellovary\altaffilmark{1}, Mordecai-Mark Mac
    Low\altaffilmark{1,2}, Barry McKernan\altaffilmark{1,3,4,5},
    K. E. Saavik Ford\altaffilmark{1,3,4,5}}

\altaffiltext{1}{Department of Astrophysics, American Museum of
  Natural History, Central Park West at 79th Street, New York, NY
  10024, USA} 
\altaffiltext{2}{Institut f\"ur Theoretische Astrophysik, Zentrum
  f\"ur Astronomie der Universit\"at Heidelberg, 69120 Heidelberg, Germany}
\altaffiltext{3}{Department of Science, Borough of Manhattan Community College, City University of New York, New York, NY 10007}
\altaffiltext{4}{Physics Program, The Graduate Center, CUNY, New York, NY 10016}
\altaffiltext{5}{Kavli Insitute for Theoretical Physics, University of
  California, Santa Barbara, Santa Barbara, CA 93106}

\begin{abstract}
  Accretion disks around supermassive black holes (SMBHs) in active
  galactic nuclei contain stars, stellar mass black holes, and other
  stellar remnants, which perturb the disk gas
  gravitationally. The resulting density perturbations exert
  torques on the embedded masses causing them to migrate through the
  disk in a manner analogous to planets in
  protoplanetary disks.  We determine the strength and direction of
  these torques using an empirical analytic description dependent on
  local disk gradients, applied to two different analytic,
  steady-state disk models of SMBH accretion disks. We find that there
  are radii in such disks where the gas torque changes sign, trapping
  migrating objects.  Our analysis shows that major migration traps
  generally occur where the disk surface density gradient changes sign
  from positive to negative, around 20--300$R_{\rm g}$, where $R_{\rm
    g}=2GM/c^{2}$ is the Schwarzschild radius.  At these traps,
  massive objects in the AGN disk can accumulate, collide, scatter,
  and accrete.  Intermediate mass black hole formation is likely in
  these disk locations, which may lead to preferential gap and cavity
  creation at these radii.  Our model thus has significant
  implications for SMBH growth as well as gravitational wave source
  populations. 
\end{abstract}

\keywords{black hole physics --- accretion disks --- galaxies:nuclei }

\section{Introduction}
At present, the observational evidence for intermediate mass black
holes (IMBHs; $M\sim 10^2$--$10^6$~M$_{\odot}$) is much less
compelling than that for supermassive black 
holes (SMBHs; $M>10^{6}$~M$_{\odot}$) or stellar mass black holes
($M\lesssim 40$~M$_{\odot}$). Several IMBH
candidates have been identified, including off-nuclear X-ray sources
such as HLX-1 (likely $\sim 10^3$--$10^5$~\msun)
\citep{Davis_2011,Servillat_2011,Straub_2014} and 
optical emission line sources in dwarf galaxies
\citep{Reines_2013,Moran_2014,Baldassare_2015}. IMBHs are a missing
link between stellar-mass black holes and SMBHs, and indeed are 
good candidates for the seeds of SMBHs \citep{Haiman_2001}.  IMBH
candidates are hard to confirm, although they are predicted to be
wandering throughout massive galaxy halos
\citep{Holley_Bockelmann_2010,Bellovary_2010} or lurking in dwarf
galaxies \citep{Van_Wassenhove_2010}.

An additional potential habitat for IMBHs is the accretion disks
around SMBHs in active galactic nuclei (AGN). Massive objects (stellar
remnants and stars) will exist in these disks
\citep{Syer_1991,Artymowicz1993,Levin_2007,Nayakshin_2007,McKernan_2012},
where they can collide, accrete and grow.  If a mechanism exists to
efficiently collect compact objects into an orbit where they can
collide, this mass buildup could result in the efficient formation of
IMBHs in AGN disks.  

Migration toward trapping orbits may be such a mechanism.  Objects
orbiting within differentially rotating disks exchange angular
momentum with the gas around them as they orbit, which results in a
torque, typically causing the objects to migrate.  Under the azimuthally
isothermal assumption, masses within disks were shown to migrate only
inwards \citep{Goldreich_1979,Ward_1997,Tanaka_2002}.  However,
\citet{Paardekooper_2006} found that in the more realistic case of an
adiabatic midplane, migration
can proceed outwards under some circumstances. 
\citet{Paardekooper_2010} used an extensive set of numerical
simulations to empirically define the conditions determining the sign
and strength of migration.  Locations where the torque changes sign
from positive to negative have outwardly migrating objects meeting inwardly
migrating objects in an equilibrium, zero-torque orbit, forming a
migration trap.  Such traps have been 
predicted to exist in protoplanetary disks \citep{Lyra_2010}, where
they can lead to rapid growth of giant planet cores
\citep{Horn_2012}. \citet{McKernan_2012} pointed out that, by analogy,
IMBH might be able to form efficiently and grow at super-Eddington
rates in SMBH accretion disks, if they contained migration
traps. Eventually, the resulting object may be able to 
clear a gap in the disk, which would produce a range of observational
signatures \citep{McKernan_2014}.

Here we show that simple, analytic, steady-state models of AGN
disks do indeed predict migration traps, at radii that are independent
of the SMBH mass and the mass ratio between the migrator and the
central SMBH. We further briefly discuss the importance and
observational implications of migration traps in AGN disks.

\section{Methods}

In this section we describe the torque model of
\citet{Paardekooper_2010}, and discuss its application to two
different steady-state AGN accretion disk models.

\subsection{Torque Model}\label{sect:torques}
The torque model is based on simulations performed to study the
behavior of objects in protoplanetary disks, but the physical
processes modeled are no different in optically-thick AGN accretion
disks.  We assume that the mass of the migrating object (i.e. a
stellar mass black hole) remains constant, and neglect accretion or
feedback effects on the gas. The torque model includes a linear estimate of the Lindblad (wave) torque plus a simple but nonlinear contribution from adiabatic corotation torques. It is valid for the unsaturated case, where a
temperature gradient is maintained by turbulent and viscous diffusion,
as opposed to the gradient being erased as angular momentum is
transferred between the migrating object and nearby gas.
Saturation can be neglected so long as the diffusion timescale is
short compared to the libration timescale on which the torque acts
\citep{Kley09}.

We model the torques using the analytical fits of
\citet{Paardekooper_2010} to a broad range of simulations that
included non-isothermal effects and a non-linear model of adiabatic
corotation torques.  For the locally isothermal case, the normalized
torque is
\begin{equation}
\Gamma_{\rm{iso}}/\Gamma_0 = -0.85 - \alpha - 0.9\beta,
\end{equation}
while for the purely adiabatic case the normalized torque is
\begin{equation}
\gamma \Gamma_{\rm{ad}}/\Gamma_0  = -0.85 - \alpha - 1.7\beta + 7.9\xi / \gamma.
\end{equation}
The adiabatic index $\gamma = 5/3$, and the variables $\alpha$,
$\beta$, and $\xi$ are the negative gradients of the local density,
temperature, and entropy, with values
\begin{equation}\label{eqn:torques}
\alpha = -\frac{\partial{\rm{ln} \Sigma}}{\partial{\rm{ln} r}};   \beta = -\frac{\partial{\rm{ln} T}}{\partial{\rm{ln} r}};    \xi =  \beta - (\gamma - 1)\alpha .
\end{equation}

The torques are normalized by
\begin{equation}\label{eqn:normalize}
\Gamma_0  = (q/h)^2 \Sigma r^4 \Omega^2,
\end{equation}
where $q$ is the mass ratio of the migrator to the
SMBH, $h$ is the aspect ratio of the disk, and $\Omega$ is the
rotational velocity.  Interpolating between the isothermal and
adiabatic torque regimes, we obtain
\begin{equation}\label{eqn:torquetotal}
\Gamma = \frac{\Gamma_{\rm ad}\Theta^2 + \Gamma_{\rm iso}}{(\Theta + 1)^2}
\end{equation}
where $\Theta$ is the ratio of the radiative and dynamical timescales
$t_{rad}/t_{dyn}$.  \citet{Lyra_2010} show that $\Theta$ depends on
the local disk properties as
\begin{equation}\label{eqn:theta}
\Theta = \frac{c_v \Sigma \Omega \tau_{\rm{eff}}}{12 \pi \sigma T^3}
\end{equation}
where c$_v$ is the thermodynamic constant with constant volume,
$\tau_{\rm{eff}}$ is the effective optical depth, and $\sigma$ is the
Stefan-Boltzmann constant.  The value of $\tau_{\rm{eff}}$ is taken at
the midplane \citep{Hubeny_1990,Kley_2008} as
\begin{equation}
\tau_{\rm{eff}} = \frac{3\tau}{8} + \frac{\sqrt{3}}{4} + \frac{1}{4\tau}
\end{equation}
where $\tau$ is the true optical depth, calculated by $\tau =
\kappa\Sigma/2$, where $\kappa$ is the opacity.  

In summary, each of the torque components depends on the properties of the temperature, density, and entropy gradients.  For particular values of these gradients, the torques may cancel, resulting in a region with zero torque, i.e. a migration trap.  Our goal is to investigate whether stable traps exist, i.e. whether there are regions where the gradient of the torque is negative.

AGN disks are sufficiently ionized for magnetorotational instability to drive turbulence. The resulting density perturbations produce stochastic torques that can drive diffusive, random walk, migration \citep{Nelson05}. \citet{Johnson06} quantify when diffusive migration dominates over advective (type I) migration. Simulations of fully-ionized regions of stratified protoplanetary disks suggest that for interesting ranges of migrator mass and radius, type I migration prevails \citep{Yang09}. Such stochastic perturbations were shown by \citet{Horn_2012} to be necessary for multiple objects to reach equilibrium orbits and collide.  We defer numerical simulations of the AGN case to future work.

\subsection{Disk Models}
We examine the torques expected in disks described by two
steady-state, analytic SMBH accretion disk models derived by
\citet[][hereafter SG]{Sirko_2003} and \citet[][hereafter
TQM]{Thompson_2005}.  These models are derived from different basic
assumptions, but both contain many characteristics we expect in
realistic AGN disks. Neither model includes direct modeling of magnetic fields, nor effects due to general relativity.

SG assume a classical thin Keplerian $\alpha$-disk
\citep{Shakura_1973} in a steady state with a constant, high,
accretion rate (Eddington ratio of 0.5).  In order to remain stable
and prevent fragmentation (i.e.\ maintain $Q \gtrsim
1$), SG assume that stars form in the outer disk.
Energetic feedback from the newly formed stars increases the velocity
dispersion and sound speed of the gas, maintaining Q close to unity,
supporting the disk against global gravitational instability and
inhibiting further star formation.  This approach is supported by the
existence of nuclear star clusters in the vicinity of SMBHs, which may
have formed in this way
\citep{Nayakshin_2006,Levin_2007,Chang_2007}. The disk opacity model
of SG is based on \citet{Iglesias_1996} for high temperatures ($T
\gtrsim 10^4$ K) and \citet{Alexander_1994} for lower temperatures.

\begin{figure}[htb!]
\begin{center}
\includegraphics[width=0.85\columnwidth]{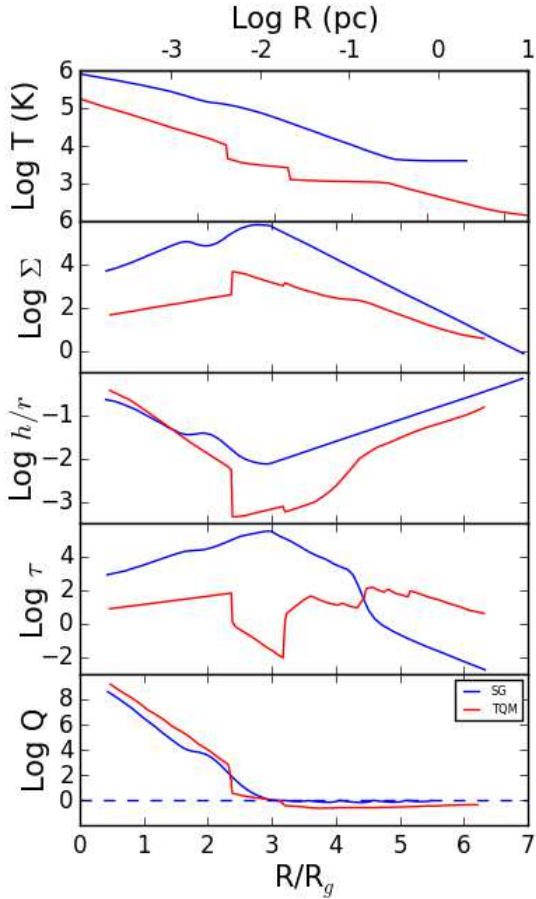}
\caption{Models of accretion disks from SG (blue) and TQM (red).  From top to bottom, we show temperature, surface density (in g cm$^{-2}$), disk aspect ratio $h/r$, optical depth $\tau$, and Toomre Q vs radius.  The top axis represents the translation from gravitational radius to parsecs for a $10^8$~\msun~SMBH.
\label{fig:profiles}
}
\end{center}
\end{figure}

The model of TQM, on the other hand, extrapolates a star-forming galaxy
disk inward to the SMBH.
Angular momentum transport is assumed to 
take place due to global gravitational instabilities, such as bars and spiral
inflows, rather than unresolved turbulent viscosity.  TQM use a more
up-to-date opacity model based on \citet{Semenov_2003}.  TQM address
gravitational fragmentation by considering two regimes: one where the external
accretion rate is high enough that the gas fraction of the disk
remains constant, allowing rapid inflow to continue; and another where
the star formation timescale is shorter than the gas advection time,
and thus accretion to the inner regions is more limited, as the gas is
consumed in star formation.

Figure \ref{fig:profiles} shows profiles from both models
of the disk temperature $T$, surface density $\Sigma$, aspect ratio $h/r$, and optical depth
$\tau$.  Although the profiles are
qualitatively comparable, there are major differences
between the models. For example, the surface
density and optical depth in SG are 2--3 orders of magnitude above
those of TQM in the inner disk. The differences in opacity and the assumed
dynamics of the inflow are 
the root cause of these differences.  SG assume that a constant
turbulent viscosity drives the inflow;
while TQM assume the inflow speed is a constant fraction of the local
sound speed. In both cases the high Thompson scattering opacity
from electrons produced by the ionization of hydrogen causes the inner
disk to be optically thick.  At intermediate radii, where the electron
density drops precipitously, the opacity drops correspondingly,
allowing the disk to cool and become thinner.  At larger radii,
where the temperature is low enough for dust grains to survive, dust opacity becomes important in the
disk, so the disk again thickens and cools further.  The disk masses (integrated out to 1 pc) are $3.7 \times 10^7$ \msun~ and $6.5 \times 10^6$ \msun~ for SG and TQM, respectively.

\section{Results}\label{sect:results}
\begin{figure}[]
\includegraphics[width=1\columnwidth]{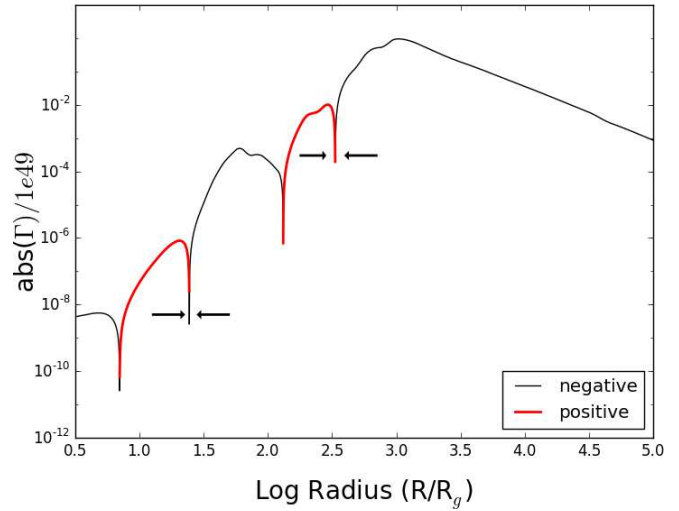}
\caption{ \label{fig:torques}
  The absolute value of the torque $\Gamma$ for the SG model, scaled
  by a factor of $10^{49}$~g~cm$^2$~s$^{-2}$, vs. gravitational radius
  R$_{\rm g}$.  Black lines indicate where the torque is negative, and
  thick red lines where it is positive.  The arrows point in the
  direction of the torque, and show that inward- and outward-pointing
  torques meet at two of the zero-crossings, forming migration traps.
}
\end{figure}

Figure \ref{fig:torques} shows the result of calculating the torques
from Equation (\ref{eqn:torquetotal}) in a disk with the profile given
by SG around a $10^8$~\msun~ SMBH for a migrator of mass 100~\msun.  The
figure shows the absolute value of the torque vs radius; black lines
represent negative torque, and thick red lines represent positive
torque. The spikes mark the points where the torque crosses zero.  The
direction of the torque is also given by arrows for clarity. These
highlight the two migration traps in this disk model: one at $\log R =
1.39 R_{\rm g}$, and the other at $\log R = 2.52 R_{\rm g}$,
corresponding to 24.5 and 331 $R_{\rm g}$, or 0.0004 and 0.003 pc for
a 10$^8$~\msun~ SMBH.  The Toomre Q parameters at the trap locations are $\sim 10^5$ and 16, respectively, indicating that these regions are quite stable.

These estimates are for a fiducial value of $M_{\rm SMBH} =
10^8$~\msun~ and mass ratio $q = 10^{-6}$. 
However, we repeated
our calculations for a range of each value ($5 \times 10^5 < M_{\rm
  SMBH} < 5 \times 10^9$~\msun~ and $0.1 < q < 10^{-6}$) and found no
difference in the radial location of the migration traps in terms of
$R_g$. We should expect this result, since the variables that depend
on the mass ratio $q$ and $M_{\rm SMBH}$ are $\Gamma_0$ and $\Theta$,
as seen in equations~(\ref{eqn:normalize}) and~(\ref{eqn:theta}). These mass adjustments change the magnitude of the torques but not their radial
position; i.e. the trap locations are not affected.  However, the SG model assumes a particular value of $M_{\rm
  SMBH}$. As we do not have access to their full set of models, we are
unable to vary the black hole mass self-consistently in our
calculations.

\begin{figure}[htb!]
\includegraphics[width=1\columnwidth]{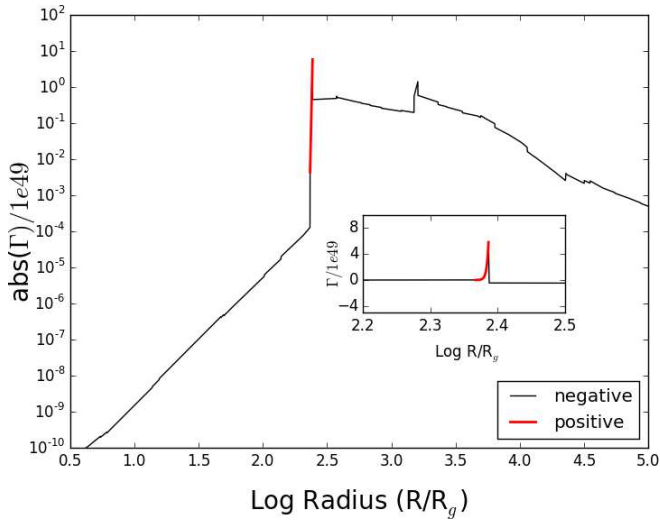}
\caption{  The absolute value of the torque $\Gamma$ for the TQM model, scaled by a factor of $10^{49}$~g~cm$^2$~s$^{-2}$, vs. normalized radius $R/R_{\rm g}$.  Black lines indicate where the torque is negative, and red thick lines where it is positive.  The inset shows $\Gamma$ on a linear scale for a small region to better visualize the migration trap.
\label{fig:thompson}
}
\end{figure}

Figure \ref{fig:thompson} shows the results for the same torque
calculation using the TQM model.  We find one migration trap, at $\log
R = 2.39 R_{\rm g}$ (245~$R_{\rm g}$, or~0.002 pc for a
10$^8$~\msun~ SMBH).  At this radius, Q = 3.5. This trap occurs precisely at the point where the
disk profiles are vertical, and the derivative is undefined (see
Figure \ref{fig:profiles}).  To explore the robustness of this result,
we made the profile differentiable by shifting the endpoints of each
vertical section of the profile to vary the slope.  In the extreme
case, we adjusted the surface density profile to effectively round off
the sharp peak at $\log R = 2.4 R_{\rm g}$.  Regardless of these
changes, the migration trap continues to exist at the point where the
surface density slope changes from positive to negative.
Significantly, migration traps also exist in the SG model at the same
locations---the points where the slope shifts from positive to
negative, indicating that the slope change of the surface density
profile is a key factor in determining where migration traps exist in
these models (see also \citet{Masset06}).

Note that in Figure~\ref{fig:profiles} there is a small surface
density discontinuity at $\log R \sim 3.2 R_{\rm g}$; however it does
not yield a migration trap in Figure~\ref{fig:thompson}.  Again we
adjusted the endpoints of the vertical section of the profile to
verify the robustness of this result.  We found that the magnitude of
the vertical change in the profile was insufficient to cause the
torque to change sign.
Thus, both a slope change {\em and} a large change in magnitude of the
surface density of an AGN disk appear to be needed in order to create
a migration trap.

\section{Implications}
The occurrence of migration traps in simple models of AGN disks
implies that IMBH may form efficiently and quickly due to stellar
black hole collisions at such locations, by analogy with giant planet
core formation at migration traps in protoplanetary disks
\citep{Horn_2012}.  Ignoring migration traps, \citet{McKernan_2012} conservatively predict that a $10 M_{\odot}$ black hole around a $10^8 M_{\odot}$ SMBH can double its mass via collisions and gas accretion in 10 Myr.  However, including migration traps can boost the collision rate of disk objects by more than a factor of 100.  For a migrator at $10^4 R_{\rm g}$ in a migration trap with enhanced surface density of compact objects of $\Sigma_0 = 350$ g cm$^{-2}$, assuming a reasonable distribution of eccentricities, the growth rate can reach over $dM/dt \sim 10^{-5} M_{\odot}$ yr$^{-1}$, which would result in a $10 M_{\odot}$ black hole growing to $\sim 100 M_{\odot}$ in 10 Myr.  We also point out that the build-up of the IMBH (i.e. the merging of stellar mass black holes) is detectable in the local universe with LIGO.  Nearby quasars such as Mrk 231 are good candidates to model and search for such events.   

If this predicted growth occurs, there are observable implications, both in electromagnetic and gravitational radiation.  For example,  if the IMBH to SMBH mass 
ratio becomes large enough ($q \geq 10^{-4}$) a gap can form in the disk at the
migration trap radius, leading to a flux decrement in the optical/UV disk
SED. \citep{Tanaka_2011,Gultekin_2012,McKernan_2014}.  

If the IMBH migrates into the central SMBH, a robust gravitational wave signal could be detected.  Such a scenario is more likely for a lower mass ($M <10^{7.5}M_{\odot}$) primary or closer-in ($200$ $R_{\rm g}$) migrator.   A binary system of mass $M_{b}=M_{1}+M_{2}$ will decay via
gravitational wave emission on a timescale \citep{Peters_1964} 
\begin{equation}
  \tau_{\rm GW} \approx \frac{5}{128}
  \frac{c^5}{G^3}\frac{a_b^4}{M_b^2\mu_b}(1-e_b^2)^{7/2},
\end{equation}

where the binary reduced mass $\mu_{b}=M_{1}M_{2}/M_{b}$, the binary
semi-major axis is $a_{b}$, and its eccentricity $e_{b}\approx 0$.
Rewriting in terms of $M_1$,$M_2$, $a_b$ and normalized by
 $R_{g1}=2GM_{1}/c^{2}$, the gravitational radius of the primary, yields
 
\begin{equation}
 \tau_{\rm GW} \approx 0.01 \rm{Myr} \left(\frac{M_1}{10^6 M_{\odot}}\right)^2 \left(\frac{M_2}{10^3 M_{\odot}}\right)^{-1} \left( \frac{a_b}{200R_{g1}}\right)^4.
\end{equation}

For a fiducial AGN disk lifetime of $\sim$10 Myr, an IMBH formed at
$200$ $R_{\rm g}$ in a disk around a SMBH with $M <10^{7.5}M_{\odot}$
should merge with the primary within the disk lifetime. If such
mergers are common, detectable gravitational wave events 
will be more frequent than previously supposed
\citep[e.g.][]{Babak_2008} and can be observed by the planned $LISA$
mission \citep[see also][]{Holley_Bockelmann_2010}, with a
complementary electromagnetic counterpart observable via oscillations
in the FeK$\alpha$ line \citep{McKernan_2013,McKernan_2015}. 

On the other hand, IMBHs that form around more massive SMBHs, or more
than twice as far away in the disk will outlast the AGN disk and
survive, potentially until the next accretion episode.  Such a binary
system can affect the galactic bulge, scattering stars and altering
the potential well. The IMBH can also itself grow due to gas accretion
\citep{Farris_2014}, changing the mass ratio of the system and
possibly being visible as a mini-quasar with shifting radial
velocities.  We will return to some of these consequences in future
work.

\section{Summary}

Migration traps are equilibrium orbits in disks where regions of outward migration meet regions of inward migration. We study migration of massive objects in AGN accretion disks to determine whether migration traps exist in such environments.  We
examine two different steady-state, analytic models of AGN disks, and
find that, despite the different assumptions used, migration traps
occur in both models.  These migration traps occur at locations
of significant change in both the magnitude and gradient of the surface density.
In the traps, massive objects, such as stellar mass black holes, can
accumulate and merge, resulting in the formation of IMBHs.
These IMBHs could ultimately clear
out a gap in the accretion disk, producing multiple observable
signatures \citep{McKernan_2014}. The buildup of IMBHs could be a significant gravitational wave source for LIGO, and mergers of these IMBHs with their
central SMBHs 
would increase the number and strain amplitude of expected
gravitational wave sources detectable by $eLISA$.

Our prediction is based on analytical models that neglect evolution
and make strong simplifying assumptions about the dynamics.  Further
studies may need to include effects shown to be important in
the protoplanetary context, including torques due to magnetic fields
\citep{Guilet_2013}, and accretion heating feedback from the migrator
\citep{Ben_tez_Llambay_2015}.  Accretion disk dynamics are more
complex than the assumptions of either SG or TQM, as can be
seen from the substantial differences between the models.
Ultimately, migration depends on the detailed physical state of
the disk, including the temperature, density, opacity, and turbulence.
We therefore stress that our results should not be interpreted
literally, but rather as a promising possibility worthy of further
detailed modeling.  We also assume that trapped compact objects will have common orbits with low eccentricity, and merge without any dynamical consequences.  Further studies must determine whether collisions of migrators will perturb the disk and affect migration, and whether the scattering of compact objects will result in ejections from the disk and prevent our scenario entirely.  A full, three-dimensional, time-evolving model
will ultimately be needed in order to make robust predictions of
whether AGN disks can efficiently form IMBHs within migration traps.

Thanks to Alex Hubbard, Yuri Levin, Cole Miller, Shane Davis, and the anonymous referee for
useful discussions. JMB acknowledges generous support from the Helen Gurley
Brown Trust.  M-MML acknowledges support from
NSF grant AST-1109395 and the Alexander von Humboldt Foundation. 
BM and KESF acknowledge support from NSF PAARE grant
AST-1153335 and NSF grant PHY-1125915.


\begin{thebibliography}{48}
\expandafter\ifx\csname natexlab\endcsname\relax\def\natexlab#1{#1}\fi

\bibitem[{Alexander \& Ferguson(1994)}]{Alexander_1994}
Alexander, D.~R., \& Ferguson, J.~W. 1994, ApJ, 437, 879

\bibitem[{{Artymowicz} {et~al.}(1993){Artymowicz}, {Lin}, \&
  {Wampler}}]{Artymowicz1993}
{Artymowicz}, P., {Lin}, D.~N.~C., \& {Wampler}, E.~J. 1993, ApJ, 409, 592

\bibitem[{Babak {et~al.}(2008)Babak, Baker, Benacquista, Cornish, Crowder,
  Larson, Plagnol, Porter, Vallisneri, Vecchio, Arnaud, Barack, B{\l}aut,
  Cutler, Fairhurst, Gair, Gong, Harry, Khurana, Kr{\'{o}}lak, Mandel, Prix,
  Sathyaprakash, Savov, Shang, Trias, Veitch, Wang, Wen, \&
  Whelan}]{Babak_2008}
Babak, S. {et~al.} 2008, Class. Quantum Grav., 25, 184026

\bibitem[{Baldassare {et~al.}(2015)Baldassare, Reines, Gallo, \&
  Greene}]{Baldassare_2015}
Baldassare, V.~F., Reines, A.~E., Gallo, E., \& Greene, J.~E. 2015, ApJ, 809,
  L14

\bibitem[{Bellovary {et~al.}(2010)Bellovary, Governato, Quinn, Wadsley, Shen,
  \& Volonteri}]{Bellovary_2010}
Bellovary, J.~M., Governato, F., Quinn, T.~R., Wadsley, J., Shen, S., \&
  Volonteri, M. 2010, ApJ, 721, L148

\bibitem[{Ben{\'{\i}}tez-Llambay {et~al.}(2015)Ben{\'{\i}}tez-Llambay, Masset,
  Koenigsberger, \& Szul{\'{a}}gyi}]{Ben_tez_Llambay_2015}
Ben{\'{\i}}tez-Llambay, P., Masset, F., Koenigsberger, G., \& Szul{\'{a}}gyi,
  J. 2015, Nature, 520, 63

\bibitem[{Chang {et~al.}(2007)Chang, Murray-Clay, Chiang, \&
  Quataert}]{Chang_2007}
Chang, P., Murray-Clay, R., Chiang, E., \& Quataert, E. 2007, ApJ, 668, 236

\bibitem[{Davis {et~al.}(2011)Davis, Narayan, Zhu, Barret, Farrell, Godet,
  Servillat, \& Webb}]{Davis_2011}
Davis, S.~W., Narayan, R., Zhu, Y., Barret, D., Farrell, S.~A., Godet, O.,
  Servillat, M., \& Webb, N.~A. 2011, ApJ, 734, 111

\bibitem[{Farris {et~al.}(2014)Farris, Duffell, MacFadyen, \&
  Haiman}]{Farris_2014}
Farris, B.~D., Duffell, P., MacFadyen, A.~I., \& Haiman, Z. 2014, ApJ, 783, 134

\bibitem[{Goldreich \& Tremaine(1979)}]{Goldreich_1979}
Goldreich, P., \& Tremaine, S. 1979, ApJ, 233, 857

\bibitem[{Guilet {et~al.}(2013)Guilet, Baruteau, \& Papaloizou}]{Guilet_2013}
Guilet, J., Baruteau, C., \& Papaloizou, J. C.~B. 2013, MNRAS, 430, 1764

\bibitem[{{G{\"u}ltekin} \& {Miller}(2012)}]{Gultekin_2012}
{G{\"u}ltekin}, K., \& {Miller}, J.~M. 2012, \apj, 761, 90

\bibitem[{Haiman \& Loeb(2001)}]{Haiman_2001}
Haiman, Z., \& Loeb, A. 2001, ApJ, 552, 459

\bibitem[{Holley-Bockelmann {et~al.}(2010)Holley-Bockelmann, Micic, Sigurdsson,
  \& Rubbo}]{Holley_Bockelmann_2010}
Holley-Bockelmann, K., Micic, M., Sigurdsson, S., \& Rubbo, L.~J. 2010, ApJ,
  713, 1016

\bibitem[{Horn {et~al.}(2012)Horn, Lyra, Low, \& S{\'{a}}ndor}]{Horn_2012}
Horn, B., Lyra, W., Low, M.-M.~M., \& S{\'{a}}ndor, Z. 2012, ApJ, 750, 34

\bibitem[{Hubeny(1990)}]{Hubeny_1990}
Hubeny, I. 1990, ApJ, 351, 632

\bibitem[{Iglesias \& Rogers(1996)}]{Iglesias_1996}
Iglesias, C.~A., \& Rogers, F.~J. 1996, ApJ, 464, 943

\bibitem[{{Johnson} {et~al.}(2006){Johnson}, {Goodman}, \& {Menou}}]{Johnson06}
{Johnson}, E.~T., {Goodman}, J., \& {Menou}, K. 2006, \apj, 647, 1413

\bibitem[{{Kley} {et~al.}(2009){Kley}, {Bitsch}, \& {Klahr}}]{Kley09}
{Kley}, W., {Bitsch}, B., \& {Klahr}, H. 2009, \aap, 506, 971

\bibitem[{Kley \& Crida(2008)}]{Kley_2008}
Kley, W., \& Crida, A. 2008, Astronomy and Astrophysics, 487, L9

\bibitem[{Levin(2007)}]{Levin_2007}
Levin, Y. 2007, MNRAS, 374, 515

\bibitem[{Lyra {et~al.}(2010)Lyra, Paardekooper, \& Low}]{Lyra_2010}
Lyra, W., Paardekooper, S.-J., \& Low, M.-M.~M. 2010, ApJ, 715, L68

\bibitem[{{Masset} {et~al.}(2006){Masset}, {Morbidelli}, {Crida}, \&
  {Ferreira}}]{Masset06}
{Masset}, F.~S., {Morbidelli}, A., {Crida}, A., \& {Ferreira}, J. 2006, \apj,
  642, 478

\bibitem[{McKernan \& Ford(2015)}]{McKernan_2015}
McKernan, B., \& Ford, K. E.~S. 2015, MNRAS: Letters, 452, L1

\bibitem[{McKernan {et~al.}(2013)McKernan, Ford, Kocsis, \&
  Haiman}]{McKernan_2013}
McKernan, B., Ford, K. E.~S., Kocsis, B., \& Haiman, Z. 2013, MNRAS, 432, 1468

\bibitem[{McKernan {et~al.}(2014)McKernan, Ford, Kocsis, Lyra, \&
  Winter}]{McKernan_2014}
McKernan, B., Ford, K. E.~S., Kocsis, B., Lyra, W., \& Winter, L.~M. 2014,
  MNRAS, 441, 900

\bibitem[{McKernan {et~al.}(2012)McKernan, Ford, Lyra, \&
  Perets}]{McKernan_2012}
McKernan, B., Ford, K. E.~S., Lyra, W., \& Perets, H.~B. 2012, MNRAS, 425, 460

\bibitem[{Moran {et~al.}(2014)Moran, Shahinyan, Sugarman, V{\'{e}}lez, \&
  Eracleous}]{Moran_2014}
Moran, E.~C., Shahinyan, K., Sugarman, H.~R., V{\'{e}}lez, D.~O., \& Eracleous,
  M. 2014, The Astronomical Journal, 148, 136

\bibitem[{Nayakshin(2006)}]{Nayakshin_2006}
Nayakshin, S. 2006, MNRAS, 372, 143

\bibitem[{Nayakshin \& Sunyaev(2007)}]{Nayakshin_2007}
Nayakshin, S., \& Sunyaev, R. 2007, MNRAS, 377, 1647

\bibitem[{{Nelson}(2005)}]{Nelson05}
{Nelson}, R.~P. 2005, \aap, 443, 1067

\bibitem[{Paardekooper {et~al.}(2010)Paardekooper, Baruteau, Crida, \&
  Kley}]{Paardekooper_2010}
Paardekooper, S.-J., Baruteau, C., Crida, A., \& Kley, W. 2010, MNRAS, 401,
  1950

\bibitem[{{Paardekooper} \& {Mellema}(2006)}]{Paardekooper_2006}
{Paardekooper}, S.-J., \& {Mellema}, G. 2006, \aap, 459, L17

\bibitem[{Peters(1964)}]{Peters_1964}
Peters, P.~C. 1964, Phys. Rev., 136, B1224

\bibitem[{Reines {et~al.}(2013)Reines, Greene, \& Geha}]{Reines_2013}
Reines, A.~E., Greene, J.~E., \& Geha, M. 2013, ApJ, 775, 116

\bibitem[{Semenov {et~al.}(2003)Semenov, Henning, Helling, Ilgner, \&
  Sedlmayr}]{Semenov_2003}
Semenov, D., Henning, T., Helling, C., Ilgner, M., \& Sedlmayr, E. 2003,
  Astronomy and Astrophysics, 410, 611

\bibitem[{Servillat {et~al.}(2011)Servillat, Farrell, Lin, Godet, Barret, \&
  Webb}]{Servillat_2011}
Servillat, M., Farrell, S.~A., Lin, D., Godet, O., Barret, D., \& Webb, N.~A.
  2011, ApJ, 743, 6

\bibitem[{Shakura \& Sunyaev(1973)}]{Shakura_1973}
Shakura, N.~I., \& Sunyaev, R.~A. 1973, in X- and Gamma-Ray Astronomy (Springer
  Science + Business Media), 155--164

\bibitem[{Sirko \& Goodman(2003)}]{Sirko_2003}
Sirko, E., \& Goodman, J. 2003, MNRAS, 341, 501

\bibitem[{Straub {et~al.}(2014)Straub, Godet, Webb, Servillat, \&
  Barret}]{Straub_2014}
Straub, O., Godet, O., Webb, N., Servillat, M., \& Barret, D. 2014, Astronomy
  {\&} Astrophysics, 569, A116

\bibitem[{Syer {et~al.}(1991)Syer, Clarke, \& Rees}]{Syer_1991}
Syer, D., Clarke, C.~J., \& Rees, M.~J. 1991, MNRAS, 250, 505

\bibitem[{Tanaka {et~al.}(2002)Tanaka, Takeuchi, \& Ward}]{Tanaka_2002}
Tanaka, H., Takeuchi, T., \& Ward, W.~R. 2002, ApJ, 565, 1257

\bibitem[{Tanaka {et~al.}(2011)Tanaka, Menou, \& Haiman}]{Tanaka_2011}
Tanaka, T., Menou, K., \& Haiman, Z. 2011, MNRAS, 420, 705

\bibitem[{Thompson {et~al.}(2005)Thompson, Quataert, \& Murray}]{Thompson_2005}
Thompson, T.~A., Quataert, E., \& Murray, N. 2005, ApJ, 630, 167

\bibitem[{Ward(1997)}]{Ward_1997}
Ward, W. 1997, Icarus, 126, 261

\bibitem[{Wassenhove {et~al.}(2010)Wassenhove, Volonteri, Walker, \&
  Gair}]{Van_Wassenhove_2010}
Wassenhove, S.~V., Volonteri, M., Walker, M.~G., \& Gair, J.~R. 2010, MNRAS,
  408, 1139

\bibitem[{{Yang} {et~al.}(2009){Yang}, {Mac Low}, \& {Menou}}]{Yang09}
{Yang}, C.-C., {Mac Low}, M.-M., \& {Menou}, K. 2009, \apj, 707, 1233

\bibitem[{{Yang} {et~al.}(2012){Yang}, {Mac Low}, \& {Menou}}]{Yang12}
---. 2012, \apj, 748, 79

\end{thebibliography}

\end{document}